\documentclass[twocolumn,showpacs,preprintnumbers,amsmath,amssymb,a4paper,prl
,aps,nobibnotes, nofootinbib,secnumarabic,tightenlines]{revtex4-1}
\usepackage{graphicx}% Include figure files
\usepackage{dcolumn}% Align table columns on decimal point
\usepackage{bm}% bold math
\usepackage{longtable}%
\begin{document}
%\ref{}
%\preprint{}

\title{Quantum Information Processing Using Excitonic States of Quantum Dots and an Optical Cavity Manipulated by Collinear Laser Fields}

\author{I Wayan Sudiarta}  \email{wayan.sudiarta@unram.ac.id}
\affiliation{
Physics Study Program, Faculty of Mathematics and Natural Sciences\\ 
the University of Mataram, Mataram, NTB 83125, Indonesia
}

\author{Jordan Kyriakidis} \email{jordan.kyriakidis@dal.ca}
\affiliation{
Department of Physics and Atmospheric Science,\\ 
Dalhousie University, Halifax, NS B3H 3J5, Canada
}

\date{\today}

\begin{abstract}

In this paper we investigate an implementation of a quantum gate for quantum information processing in a system of quantum dots in an optical cavity manipulated by collinear laser fields. For simplicity we give theoretical and numerical results only for simulations of two quantum dots in a cavity interacting with two collinear fields. Extension to the system of many quantum dots in a cavity can be done in similar manner as the two dots system. It is shown that due to the collinear fields are used, a two qubit gate operation can be acheived by choosing properly detunings and amplitudes of the collinear fields.
\end{abstract}

%\pacs{Pacs}
\maketitle

Present day computers are based mainly on semiconductor transistors as their basic units. As the size of transistors become smaller, they will reach a size where the quantum properties of electrons can no longer be neglected. This indicates that a new device based on quantum mechanics will be needed. Moreover, it has been shown recently that quantum mechanics and with quantum algorithms, such as Shor's integer factorization and Grover's search algorithms \cite{Shor1997,Grover1997}, can offer a new way of computing that is exponentially faster than current computers.

In order to develop quantum computers, we require two basic ingredients, quantum bits (qubits) and entanglement between qubits. Beside that we need also operational gates to perform single- and two-qubit operations. It can be shown that all quantum algorithms can be constructed from a single-qubit gate and in combination with a controlled-NOT gate or a $\sqrt{\text{SWAP}}$ gate \cite{Deutsch1989, Barenco1995, Nielsen2010}. Other requirements of a quantum computer can be found in \cite{Divincenzo2001}. The qubits can be in any forms. One successful example is ions in an ion trap \cite{Cirac1995} where qubits are in the form of internal states of the ions and the entanglement between the ions is mediated by their electrostatic interaction. The communication between ions is done using their collective vibrational modes and quantum gates are performed by using lasers which interact with the ions. The first experimental implementation of a C-NOT gate in two ions system was reported by Monroe et al. \cite{Monroe1995} following a scheme given by Cirac and Zoller \cite{Cirac1995}. However, to apply the Cirac and Zoller's scheme requires the ions to be in their joint motional ground state. This implies that the scheme can be affected by the change in the vibrational modes, and consequently, the temperature of the environment.  An alternative scheme proposed by Molmer and Sorensen \cite{Sorensen1999,Molmer1999} does not have this restriction. The experimental implementations of the Molmer and Sorensen's scheme have been reported by Benhelm et al.\cite{Benhelm2008} for two ions and Sackett et al. \cite{Sackett2000} for four ions. 

Instead of ions, semiconductor quantum dots can offer more flexibility and scalability for quantum computers. The quantum dots can be constructed and their energy levels can be tuned by selecting appropriately the size of the quantum dots and the magnitude and the orientation of an external magnetic field. Qubits using the quantum dots can be based on spin states of one-electron quantum dots or the ground and excitonic states of quantum dots. The coupling between quantum dots can be done by putting the dots in an optical cavity where the entanglement is mediated by cavity photons. Gate operations are performed by manipulating the quantum dots with focused laser beams. Single-qubit operations have been proposed and reported recently \cite{Emary2007, Caillet2007, Greilich2006, Press2008, Kosaka2009}. Proposal for two-qubits operations have been reported by Imamoglu et al. \cite{Imamoglu1999} and Feng and co-workers \cite{Feng2004,Feng2002a,Feng2002b,Feng2003,Feng2005}. However, all of these proposals are for spin-based qubits. Imamoglu et al. showed that a conditional-phase-flip (CPF) gate can be performed by addressing each quantum dot with a single laser beam. This proposal is difficult to implement experimentally since the size of focused laser beam is much larger than the size of quantum dots. Conceptually this proposal is similar to the Cirac and Zoller's scheme since it requires the system be in the vacuum state of cavity. In other words gate operations work well for zero initial number of photons in the cavity. It can be shown that as the number of photons in the cavity increases the interaction between the quantum dots via cavity becomes less effective.

In this paper, we focus on qubits based on two states in the quantum dots which, in this case, are the ground state and an excitonic state. The quantum dots can be either charged or neutral quantum dots. The charged quantum dots may offer more flexibility due to additional variables that can be tuned using light polarizations and magnetic fields utilizing the Pauli-blocking mechanism\cite{Pazy2003}. Due to a short decoherence time of excitonic states, we need to perform computational processing in a short period of time. This is not a big hurdle since the pulse duration of laser can de done in about femto seconds range. 

In this paper we propose an operational gate that is similar to the Sorensen and Molmer's scheme. However, there are two main differences in our system where we use collinear laser beams for all qubits manipulations and detunings are the same for both laser fields. Since all the quantum dots are in the same laser fields, one requirement for this system to work is that the quantum dots must have different energy levels. This means that the quantum dots must be of different sizes. The energy levels can be also controlled by a magnetic field using Zeeman splitting of spin and trions states.

To simplify the analysis, we consider only two quantum dots in a cavity. Generalization for many quantum dots can be done in similar manner. Each quantum dot is modeled by a two-level system with states denoted by $|0\rangle$ and $|1\rangle$, interacting with a cavity field and two laser fields. The state $|0\rangle$ is the ground state of the quantum dot and the state $|1\rangle$ is the excitonic state. In this system the cavity and laser fields are the common fields between the two quantum dots. The Hamiltonian of the system is given by

\begin{align}
& \hat{H} =  \hbar \omega_c \hat{a}^\dagger \hat{a} + \sum_{j=1}^2 \hbar \omega_{j,0} \sigma_{j,00} + \hbar \omega_{j,1} \sigma_{j,11}+\hat{H}_{I} 
\end{align}
and the interaction term $\hat{H}_{I}$ is given by
\begin{align}
\hat{H}_{I} = -\hbar \sum_{j=1}^2 \sum_{k=1}^2 &
\frac{\Omega_{jk}}{2}  \hat{a}^\dagger \sigma_{j,01} e^{i\omega_k t} \nonumber \\
&+\frac{\Omega_{jk}}{2} \sigma_{j,01} e^{i\omega_k t}  \hat{a}
+\text{H.c.}    
\end{align}
where $\text{H.c.}$ means Hermitian conjugate, the index j denotes the index of the quantum dots and the index k denotes the index of laser fields. $\Omega$s defined by $\Omega_{jk} = g_{c,j}g_{L,j,k}/\nu_i$ is the effective Rabi frequency of jth quantum dot with respect to kth laser field and the cavity field. $g_{c,j}$ and $g_{L,j,k}$ are the coupling strength of cavity and kth laser field with jth quantum dot. $ \nu_i$ is the detuning with respect to an intermediate state. We use a notation $\sigma_{j,ab} = |a_j\rangle \langle b_j| $. $\omega_k$ is the angular frequency of the kth laser field and $\omega_{j,i}$ is the frequency corresponding to the ith energy level of the jth quantum dot. $\omega_c$ is the frequency of the cavity field. It is noted that the above Hamiltonian is similar to the Hamiltonian for ions in an ion-trap as given in \cite{Sorensen1999} when the Lamb-Dicke parameter $\eta$ is small. For the rest of this paper we set $\hbar\omega_{1,0}=\hbar\omega_{2,0} = 0$ as a zero reference energy. This means $\hbar(\omega_{1,1}-\omega_{2,1}) = \hbar\delta$ is the difference between energies of excitonic level of  first and second quantum dots.    

In a rotating frame, the Hamiltonian becomes 
\begin{align}
\hat{H} = -\hbar \sum_{j=1,2} \sum_{k=1,2} & 
\frac{\Omega_{jk}}{2}  \hat{a}^\dagger \sigma_{j,01} e^{i(\Delta_{jk}+\omega_c) t} \nonumber \\
&+\frac{\Omega_{jk}}{2} \sigma_{j,01} e^{i(\Delta_{jk} - \omega_c) t}  \hat{a}
+\text{H.c.}    
\label{eqn-hrot}
\end{align}
where the detunings are defined by $\Delta_{jk} = \omega_k - (\omega_{j,1} - \omega_{j,0})$.

To simplify the analysis, let us first assumed that each quantum dot interacts with only one laser field. This can be achieved by selecting frequencies such that $(\omega_1 -\omega_2)$ is much larger than the coupling constants $\Omega_{jk}$ and $(\Delta_{jk}-\omega_c)$. The Hamiltonian in Eq.~(\ref{eqn-hrot}) simplifies to
\begin{align}
\hat{H} = -\hbar \sum_{j=1}^2  &
\frac{\Omega_{jj}}{2}  \hat{a}^\dagger \sigma_{j,01} e^{i(\Delta_{jj}+\omega_c) t} \nonumber \\
&+\frac{\Omega_{jj}}{2} \sigma_{j,01} e^{i(\Delta_{jj} - \omega_c) t}  \hat{a}
+\text{H.c.}
\label{eqn-hrot2}    
\end{align}

\begin{figure}[htp]
\centering
\includegraphics[width=0.5\textwidth]{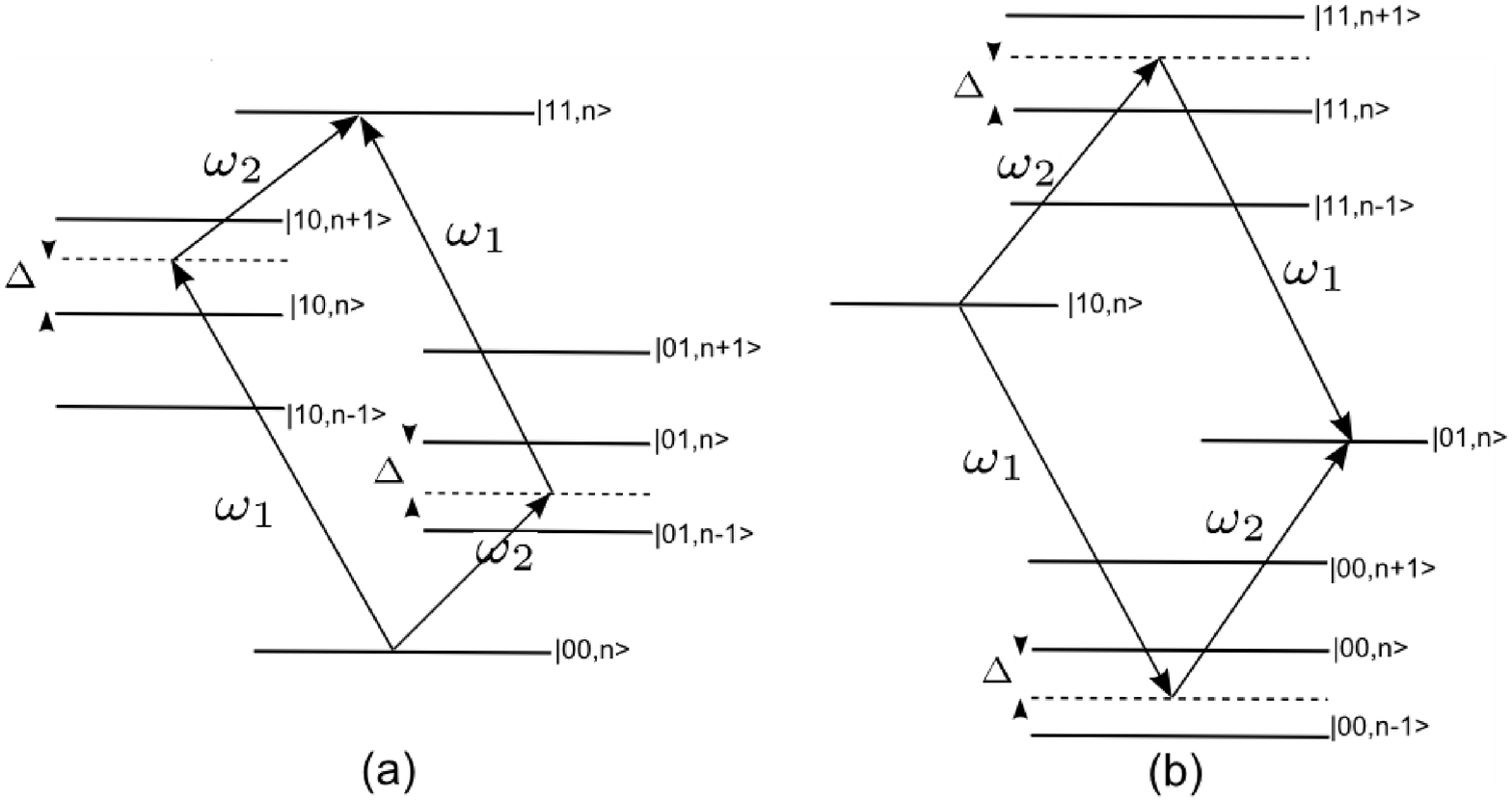}
\caption{Two configurations of two-qubit gate operations: (a) Scheme 1 and (b) Scheme 2.}
\label{fig-1}
\end{figure}

Based on this Hamiltonian, there are two possible two-qubit gate operations can be performed on the two quantum dots-cavity system. The operations are dependent on the detunings of the laser fields, $\Delta_{11}$ and $\Delta_{22}$. One possible gate (or scheme 1) is when we use the detunings $\Delta_{11} = -\Delta_{22} = \Delta$ as shown in Fig.~\ref{fig-1}(a) (as proposed in \cite{Sorensen1999,Molmer1999}). This gate operation is to produce a transition between the state $|00,n\rangle$ and the state $|11,n\rangle$. Another possible gate (or scheme 2) is to use detunings $\Delta_1 = \Delta_2 =\Delta$ as in Fig.~\ref{fig-1}(b) where this produces a transition between the state $|01,n\rangle$ and the state $|10,n\rangle$.  

For implementation of two-qubit gates using quantum dots - cavity and laser fields, it is required that all frequencies used are in optical range. It can be shown that in order for the above schemes to work properly, the detunings, $\Delta_{jj}$, must be in the same order of magnitude as the cavity frequency or $\Delta \sim \omega_c$. For the scheme 1, since $\Delta_{11} = -\Delta_{22} = \Delta$ and energy levels of quantum dots are approximately similar, it is found that the difference of the two laser fields must be in the order of $2\omega_c$. This implies that we have to use two different laser fields (one in infrared and one in blue light frequency). This may not be feasible experimentally. However, this requirement is not needed for the scheme 2. It can be shown that since $\Delta_{11} = \Delta_{22}$, the two lasers fields are in the same frequency range. This means that the scheme 2 can be implemented experimentally. For this reason in this paper we focus mainly on the scheme 2. To our knowledge, this scheme 2 has not been studied previously. 

Since $\omega_c$ is comparable to the laser frequencies, it is a good approximation to assume that in the evolution of states $|01,n\rangle$ and $|10,n\rangle$ can be described sufficiently using states $|01,n\rangle$, $|00,n-1\rangle$, $|00,n+1\rangle$, $|11,n-1\rangle$, $|11,n+1\rangle$ and $10,n\rangle$. Using these states as the basis states for solving the Schr\"odinger equation with the Hamiltonian in Eq.~(\ref{eqn-hrot2}) and after adiabatic elimination of the states $|00,n-1\rangle$, $|00,n+1\rangle$, $|11,n-1\rangle$, and $|00,n+1\rangle$, it is found that the effective Rabi frequency for $\Omega_{11} = \Omega_{22} = \Omega$ and $\Delta_{11}=\Delta_{22}=\Delta$ is 
\begin{equation}
\Omega_{eff} = \frac{\Omega^2}{2} \left[\frac{1}{\Delta+\omega_c} - \frac{1}{\Delta-\omega_c} \right]
\label{eqn-omegaeff}
\end{equation}

It can be shown from the adiabatic elimination results that the maximum probabilities of the states $|00\rangle$ and $|11\rangle$ being occupied during the evolution of states are given by
\begin{equation}
p(00) = \frac{\Omega^2}{4} \left[\frac{n+1}{(\Delta+\omega_c)^2} + \frac{n}{(\Delta-\omega_c)^2} \right]
\label{eqn-prob0}
\end{equation}
and
\begin{equation}
p(11) = \frac{\Omega^2}{4} \left[\frac{n}{(\Delta+\omega_c)^2} + \frac{n+1}{(\Delta-\omega_c)^2} \right]
\label{eqn-prob1}
\end{equation}

\begin{figure}[htp]
\centering
\includegraphics[width=0.4\textwidth]{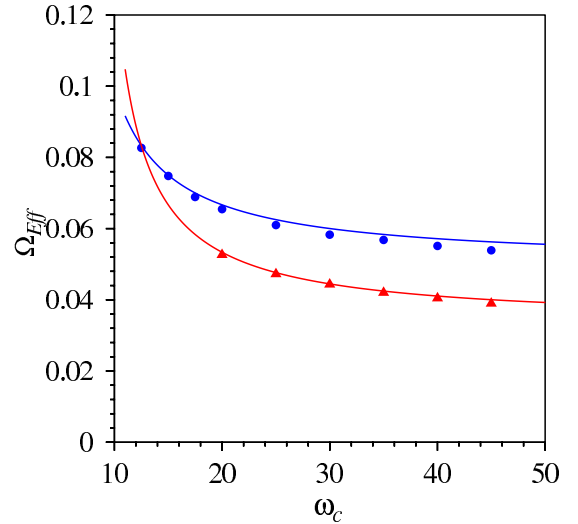}
\caption{The effective Rabi frequency as a function of $\omega_c$ with parameters $\Omega_{11}=\Omega_{22}=\Omega = 1$ and $\Delta_{11} = \Delta_{22} = \Delta$. The upper curve is for $\Delta = \omega_c - 10$ and the lower curve is for $\Delta = \omega_c - 15$. The dots and the triangles are the numerical results and the lines are computed using Eq.~(\ref{eqn-omegaeff}). }
\label{fig-2}
\end{figure}

\begin{figure}[htp]
\centering
\includegraphics[width=0.4\textwidth]{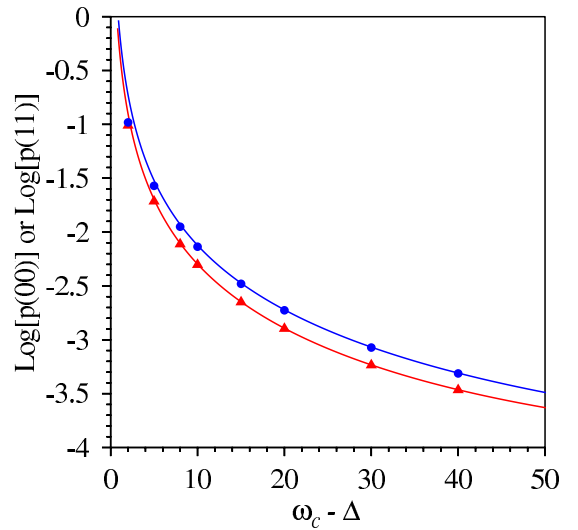}
\caption{The maximum probability of states $|00\rangle$ and $|11\rangle$ being occupied as a function of  $(\omega_c - \Delta)$ for $\omega_c = 100$.  The lower curve is for the state $|00\rangle$ and the upper curve is for the state $|11\rangle$. The triangles and dots are the numerical results for states  $|00\rangle$ and $|11\rangle$ respectively. The lines are computed using Eqs.~(\ref{eqn-prob0}) and (\ref{eqn-prob1}).}
\label{fig-3}
\end{figure}

In this paper, a dimensionless units for frequency such that $\Omega_{1} =1$ is used. Therefore, the units of time is $\Omega^{-1}$. Comparisons of Eqs.~(\ref{eqn-omegaeff}) - (\ref{eqn-prob1}) with numerical results are shown in Figs.~\ref{fig-2} and \ref{fig-3}. It is shown that the numerical results are in good agreement with the analytical results.

For cases of the two collinear fields, it is studied primarily by solving the Schr\"odinger equation numerically. In this study, a Matlab package provided by Tan \cite{Tan1999, TanUrl} is used. In all of our results we use $\Omega_{11}=\Omega_{21} =\Omega_1$ and $\Omega_{12}=\Omega_{22} =\Omega_2$. The detunings $\Delta_{11} = \Delta_{22} = \Delta$ are used. Therefore, the other detunings are $\Delta_{12} = \Delta - \delta$ and $\Delta_{12} = \Delta + \delta$. In order for this system can be extended to multi-dots system, we assume that the difference of energies between two quantum dots is $\hbar \delta = \hbar \alpha(\omega_c -\Delta)$. 

\begin{figure}[!htp]
\centering
\includegraphics[width=0.5\textwidth]{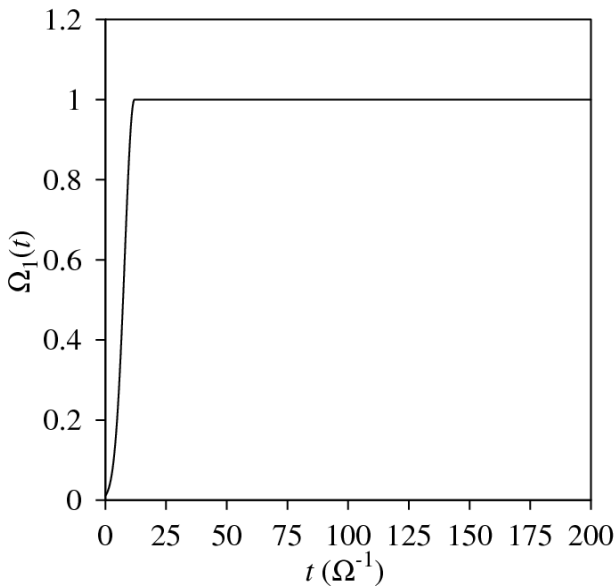}
\caption{The coupling parameter $\Omega_1(t)$ as a function of time.}
\label{fig-4}
\end{figure}

\begin{figure}[!htp]
\centering
\includegraphics[width=0.5\textwidth]{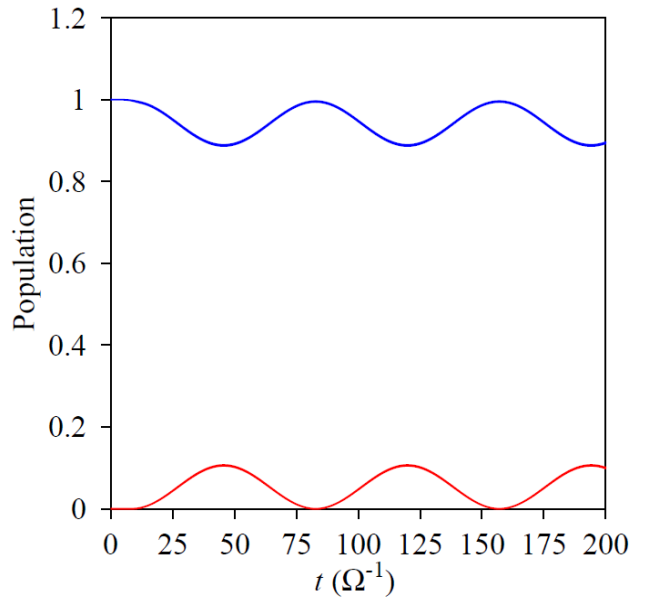}
\caption{Populations of states $|01\rangle$ and $|10\rangle$ as a function of time for parameters $\Omega_1(t) = \Omega_2(t)$ (see Fig.\ref{fig-4} for $\Omega_1(t)$), $\omega_c = 100$, $(\omega_c - \Delta) = 20$, $\alpha = 2$ and the initial state is $|01,2\rangle$. The upper (lower) curve is for the state $|01\rangle$ ($|10\rangle$). The population of states $|00\rangle$ and $|11\rangle$ are not shown here since their populations are less than 0.006.}
\label{fig-5}
\end{figure}

\begin{figure}[!htp]
\centering
\includegraphics[width=0.5\textwidth]{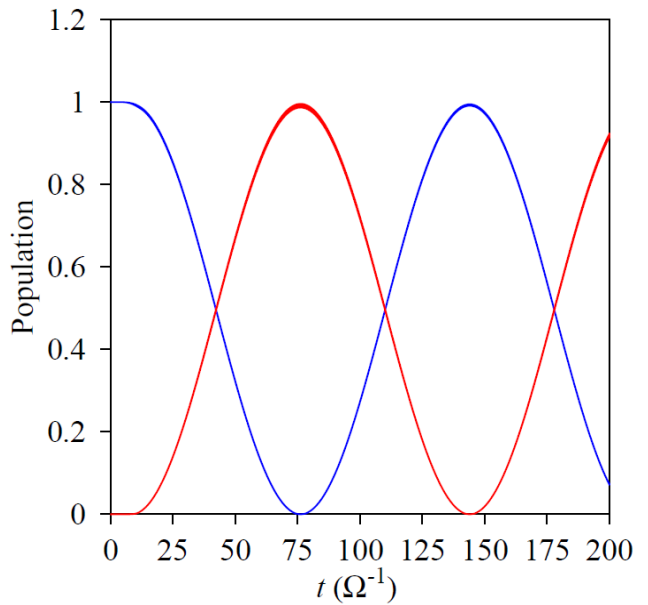}
\caption{The same as Fig.\ref{fig-5} except that this is for $\Omega_2(t) = 1.6837\Omega_1(t)$ and the population of states $|00\rangle$ and $|11\rangle$ are found to be less than 0.015. The curve with the initial value of 1 (0) is for the state $|01\rangle$ ($|10\rangle$).}
\label{fig-6}
\end{figure}

The numerical results for a continuous wave laser fields which is slowly turned on at $t=0$ for $\Omega_1(t) = \Omega_2(t)$, $\Delta_1 = \Delta_2 = 80$, $\omega_c = 100$, $\alpha = 2$ and $\langle n \rangle = 2$ are shown in Figs.~\ref{fig-4} and \ref{fig-5}. The initial state at $t=0$ is $|01,2\rangle$. It is noted that there is small amplitude of oscillation or transition between the state $|01\rangle$ and the state $|10\rangle$ contrary to our expectation when we consider the case of individual laser illumination of the quantum dots as indicated by Eq.~(\ref{eqn-omegaeff}). However, by selecting appropriately the ratio of $\Omega_2(t)$ to $\Omega_1(t)$, the transition can be improved as shown in Fig.~\ref{fig-6} using parameters $\Omega_2(t) = 1.6837\Omega_1(t)$. Using a similar derivation as in obtaining Eq.~(\ref{eqn-omegaeff}), it can be shown that the ratio of $\Omega_2$ to $\Omega_1$ needed to get a maximum amplitude of Rabi oscillation is
\begin{align}
\beta = \frac{\Omega_{2}}{\Omega_{1}} = \left[\frac{1+\frac{1}{\alpha - 1} + \frac{\Delta -\omega_c}{\Delta +\omega_c}-\frac{\Delta -\omega_c}{(\alpha + 1)\omega_c - (\alpha  - 1)\Delta}}{1-\frac{1}{\alpha + 1} + \frac{\Delta -\omega_c}{\Delta +\omega_c}-\frac{\Delta -\omega_c}{(\alpha + 1)\Delta - (\alpha -  1)\omega_c}}\right]^\frac{1}{2} 
\label{eqn-beta}
\end{align}

and the effective Rabi frequency is found to be
\begin{equation}
\Omega_{eff} = \frac{\Omega_1\Omega_2}{2} \left[\frac{1}{\Delta+\omega_c} - \frac{1}{\Delta-\omega_c} \right]
\end{equation}

It is noted that in Eq.~(\ref{eqn-beta}) that the ratio $\beta$ approaches 1 in the limit of a large $\alpha$. This is expected since for the large $\alpha$ the laser fields behave as if they are separated fields. 

It is mentioned previously that the $\sqrt{\text{SWAP}}$ gate with single gate operations are enough to perform quantum computations. The $\sqrt{\text{SWAP}}$ gate can be performed using the scheme 2 with pulsed laser fields. To confirm this gate, we compute numerically using a Gaussian shaped pulses with enveloves given by $\Omega_1(t) = \exp[-(t-t_0)^2/2\tau^2]$ and $\Omega_2(t) = \beta \exp[i\pi]\Omega_1(t)$. Note that there is a phase difference of $\pi$ between the two lasers. The parameter $\tau$ is adjusted to optimize the gate operation. The numerical results of the $\sqrt{\text{SWAP}}$ gate matrix is given in Eq.~\ref{eqn-sqrtswap} for the parameters, $\Delta_1 = \Delta_2 =990$, $\omega_c = 1000$ and $\alpha = 10$. It is found that the parameter $\tau = 16.8763$.      

\begin{equation}
\begin{pmatrix}
1 & 0 & 0 & 0 \\
0 & 0.5000 + 0.5039i & 0.5000 - 0.4961i & 0 \\
0 & 0.5000 - 0.4961i & 0.5000 + 0.5039i & 0 \\
0 & 0 & 0 & 1 \\
\end{pmatrix}
\label{eqn-sqrtswap}
\end{equation}

It is shown in Fig. \ref{fig-3} that the population of states $|00\rangle$ and $|11\rangle$ decrease as the parameter $(\omega_c - \Delta)$ increases. This means that the gate operation can be improved by increasing the value of $(\omega_c - \Delta)$ to a desired accuracy.

From Eq.~(\ref{eqn-beta}) and (\ref{eqn-omegaeff}), it may be concluded that this scheme 2 does not affected by the value of the photons number since both equations are independent of $n$. However, it is noted that Eqs.(\ref{eqn-prob0}) and (\ref{eqn-prob1}) are dependent on $n$. This indicates that in order for the $\sqrt{\text{SWAP}}$ gate operation not affected by the photons number, the parameter $\sqrt{n+1}\Omega$ should be kept much smaller than $(\omega_c-\Delta)$ during the gate operation. This condition is similar as in Molmer and Sorensen's scheme\cite{Sorensen1999}.

\bibliography{references}

\end{document}